# Dissipation Scaling and Structural Order in Turbulent Channel Flows

T.-W. Lee

Mechanical and Aerospace Engineering, SEMTE

Arizona State University, Tempe, AZ 85287

**Abstract**

Scaling and structural evolutions are contemplated in a new perspective for turbulent channel flows. The total integrated turbulence kinetic energy remains constant when normalized by the friction velocity squared, while the total dissipation increases linearly with respect to the Reynolds number. This serves as a global constraint on the turbulence structure. Motivated by the flux balances in the root turbulence variables, we also discover dissipative scaling for $u'^2$ and $v'^2$, respectively through its first and second gradients. This self-similarity allows for profile reconstructions at any Reynolds numbers based on a common template, through a simple multiplicative operation. Using these scaled variables in the Lagrangian transport equation derives the Reynolds shear stress, which in turn computes the mean velocity profile. The self-similarities along with the transport equations render possible succinct views of the turbulence dynamics and computability of the full structure in channel flows.

**INTRODUCTION**

Turbulence has been an elusive and interesting problem, as recounted in some exemplary articles (Adrian, 2010; Buschmann and Gad-el-Hak, 2003; Marusic et al., 2010). Some of the ideas developed to comprehend this phenomenon has received more attention than others, and perhaps the Reynolds number scaling is one of them. For wall-bounded flows, the scaling parameter that has been useful is the friction velocity ($u_\tau$), and related inner coordinate, y+ (Panton, 2013). For boundary-layer flows with zero pressure gradients, the normalized one-dimensional turbulence kinetic energy, $u'^{2+} = u'^2/u_\tau^2$, exhibits a progression of profiles, with the peak sharpening and moving closer to the wall with increasing Reynolds number (deGraaf and Eaton, 2000; Hutchins et al., 2009). For channel flows, the pattern is similar, but interestingly the peak location in the inner (wall) coordinates is mostly invariant at y+ ~ 15.4 (Hultmark et al., 2010). The negative peak in the Reynolds shear stress similarly moves closer to the wall with a gradual ascent toward zero centerline boundary condition, suggesting a possible dynamical connection. Scalability of these root turbulence variables will have significant implications on the structure of channel flows, including the mean velocity profiles (Hultmark et al., 2010; Barrenblatt et al., 1997).

Recently, we have derived a set of transport equations for $u'^2$, $u'v'$, and $v'^2$, based on the Lagrangian analysis of the momentum and energy transport for a control volume moving at the local mean velocity (Lee, 2020; Lee, 2021); and they provoke some ideas toward the scalability of turbulence variables. A symmetrical set of dynamical equations arose (Lee, 2020; Lee, 2021):

u′ momentum transport:

$$\frac{d(u'v')}{dy} = -C_{11} U \frac{d(u'^2)}{dy} + C_{12} U \frac{dv'^2}{dy} + C_{13} \frac{d^2 u'}{dy^2} \qquad (1)$$

$v'$ momentum transport:

$$\frac{d(v'^2)}{dy} = -C_{21} U \frac{d(u'v')}{dy} + C_{22} U \frac{dv'^2}{dy} + C_{23} \frac{d^2 v_{rms}}{dy^2} \qquad (2a)$$

Alternatively,

$$\frac{d(v'^2)}{dy} = \frac{-C_{21} U \frac{d(u'v')}{dy} + C_{23} \frac{d^2 v_{rms}}{dy^2}}{1 - C_{22} U} \qquad (2b)$$

$u'^2$ transport:

$$\frac{d(u'^3)}{dy} = -C_{31} \frac{1}{U} \frac{d(u'v' \cdot u')}{dy} + C_{32} \frac{1}{U} \frac{d(v' \cdot u'v')}{dy} + C_{33} \frac{1}{U} \left(\frac{du'}{dy}\right)^2 \qquad (3)$$

In these equations, the fluctuating terms, u'v', v'², etc., are implicitly Reynolds-averaged. The concepts and hypotheses contained in Eqs. 1-3, and their efficacy in prescribing the Reynolds stress tensor, are described in Lee (2021), and again briefly later in this work. We can see that Eqs. 1 and 2 are momentum-conserving, while Eq. 3 is an expression of energy balance. The transport equations involve gradients up to the second, as flux terms. If we compare, as an example, the Eq. 3 result with DNS data (Graham et al., 2016) for gradient of the one-dimensional turbulence kinetic energy flux (u'³), then the validation of the interior dynamics can be depicted in Figure 1. We observe that the agreement is quite close and that it is the gradient of this variable that exhibits a potentially collapsible pattern with the zero-crossing point (the u'² peak location) serving as the focal point. The gradient structure may be related to the dissipation ~ $(du'/dy)^2$. This inspires the "dissipation scaling", to examine the gradient structures in the flow. Thus envisaged, we attempt several scaling concepts that lead to some interesting self-similarity characteristics for the root turbulence variables, $u'^2$, $v'^2$, and also to some limited extent $u'v'$.

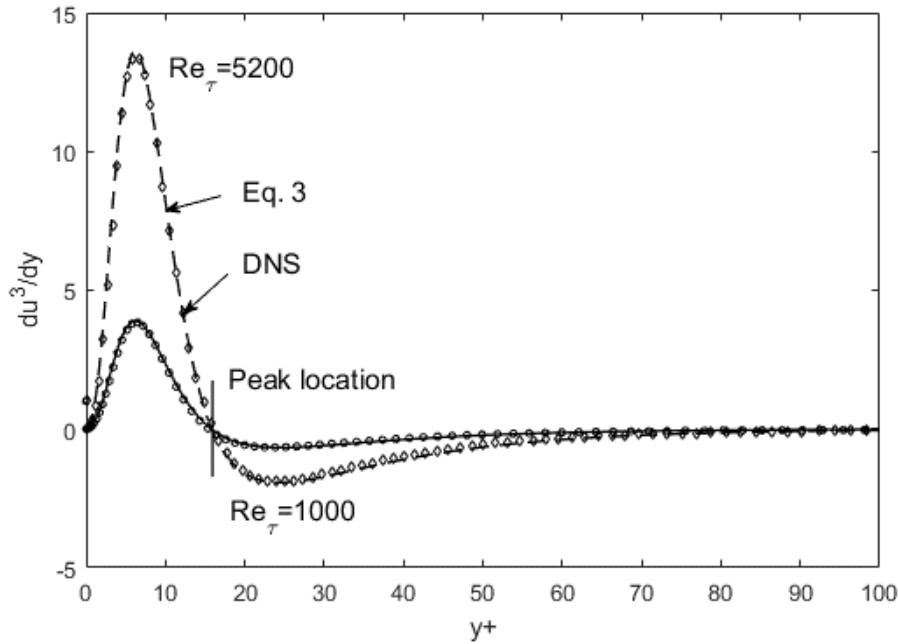

**Figure 1. Gradients of turbulence kinetic energy flux (computed using Eq. 3) are compared with DNS data (Graham et al., 2016), for $Re_\tau$=1000 and 5200. A rough picture emerges of self-similarity with a fixed peak location (zero crossing for the gradient).**

Any parameterizable scaling that may exist in the system, along with the complete set of transport (Eqs. 1-3 and the Reynolds-averaged Navier-Stokes (RANS)) equations, would establish a viable route for reconstructing the turbulence structure at any Reynolds numbers: RANS has the unknown $u'v'$, dynamics of which is prescribed by Eq. 1. In turn, this $du'v'/dy$ equation contains $v'^2$ and $u'^2$ terms, interactively governed by Eqs. 2 and 3 (Lee, 2021). Therefore, any self-similarities for these variables will have implications on the solvability of the RANS. In this paper, we would like to promulgate some of the findings on the scaling and structural characteristics in turbulent channel flows, in route to applying the theory for other more complex flow geometries (in succeeding articles).

**DISSIPATION SCALING**

Let us begin by taking a brief look at the evolution of the one-dimensional turbulence kinetic energy, $u'^2$, for a sequence of the Reynolds number ($Re_\tau$) as shown in Figure 2. The data are from DNS work of Iwamoto et al. (2002) and Graham et al. (2016). In this work, $u'^2$ and other turbulence variables are implicitly Reynolds-averaged, and also normalized by the friction velocity squared, $u_\tau^2$. For example, $<u'^{2+}> = <u'^2>/u_\tau^2$, but written as $u'^2$, to abbreviate the notation. A pronounced feature of the $u'^2$ curves is the movement of the peak location toward the wall, and it is known that when plotted in the inner coordinate ($y^+$) this point tends to stay fixed (Hultmark et al., 2010). The peak height elevates with increasing Reynolds number, even after normalizing by $u_\tau^2$ (as done in Figure 2). No single function seems capable of collapsing the curves with any kind of stretching or scaling, particularly in view of sharp peaks followed by an abrupt bend close to the wall, e.g. for $Re_\tau$=5200. However, if we estimate ("eyeball") the area under each curve in Figure 2, a possibility emerges wherein the total integrated turbulence kinetic energy (E, as defined by Eq. 4) may be constant.

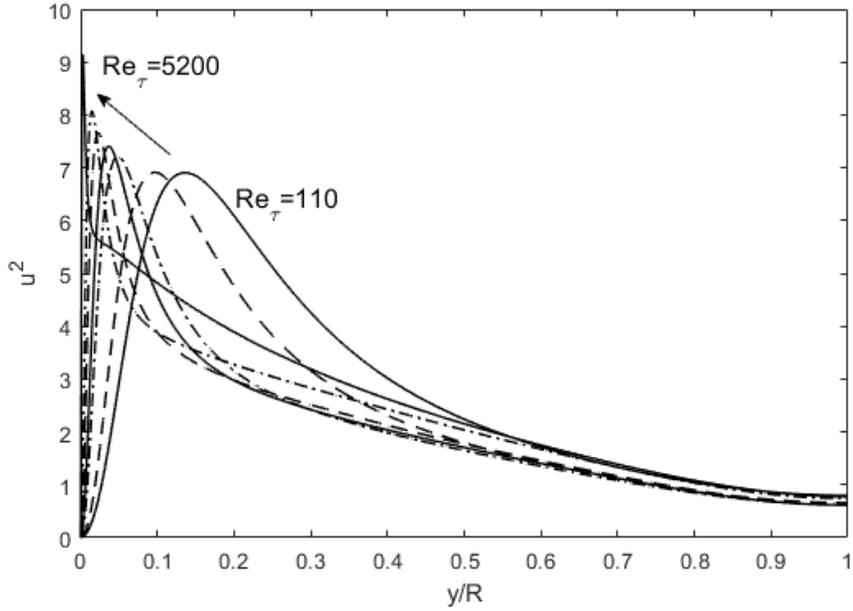

**Figure 2. The progression of u′² profiles from Re$_\tau$=110 to 5200. DNS data of Iwamoto et al. (2002) and Graham et al. (2016) are used).**

Indeed, numerical integration shows that this total energy E (Eq. 4) is invariant with respect to the Reynolds number. This owes to the fact that u$_\tau$ is a momentum scale representing the total available momentum, which would then render u$_\tau^2$ to become an energy scale. Therefore, normalizing by a global energy scale results in constant E, as shown in Figure 3.

$$E = \int_0^1 u'^{+2}(y) d\left(\frac{y}{R}\right) \tag{4}$$

$$\Phi = \int_0^1 \left(\frac{du'^+}{dy}\right)^2 d\left(\frac{y}{R}\right) \tag{5}$$

Earlier in Figure 1, we saw that the gradient of u′² may have some useful scalable characteristics. Prior to that consideration, we define (as in Eq. 5) and plot a total integrated dissipation (Φ) based on the u′² gradient, which presents a linear dependence on the Reynolds number in Figure 3. These global characteristics re-affirm $u_\tau^2$ as a useful energy scale, and also provoke the following interpretations: (1) the total turbulence kinetic energy E remains constant in spite of the "internal" transport which distributes u′² in progressively skewed curves; and (2) the dissipation (Φ) must increase with the Reynolds number as the restraining effect of the viscosity is reduced relative to the turbulence energy that exists in the flow. These may be considered as global constraints on the u′² distribution: E=const. and $\Phi = ARe_\tau$ (A is a constant). If we perceive the dissipation (skewness) through a thermodynamic lens, then it is recognizable as the degree of disorder that exists in the system. When the restoring force of viscosity is relatively reduced at higher Reynolds numbers then more dissipation must occur so that Φ proportionately increases.

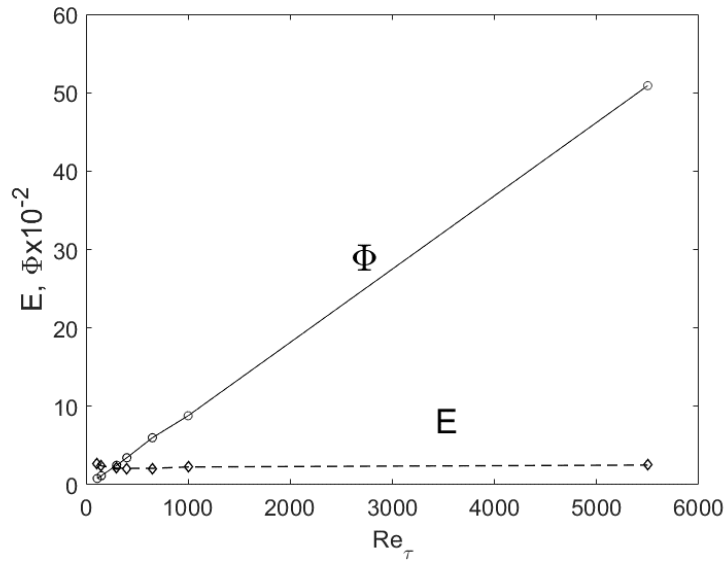

**Figure 3. The total turbulence kinetic energy (E) and dissipation (Φ) as a function of Reynolds number. DNS data are from Iwamoto et al. (2002) and Graham et al. (2016).**

The above observations and the transport equations (Eqs. 1-3) impel us to ponder upon the internal dissipation structure. Plots of $du'^2/dy$ or the "local dissipation", in $y/R$ coordinates (Figure 1) indicated some potential scaling. Recalling that an invariant $u'^2$ peak location in $y+$ coordinate is observed in experimental (Hultmark et al., 2010) and DNS (Iwamoto et al., 2002, Graham et al., 2016) data, we re-define a local dissipation in $y+$ coordinate space, as $\varepsilon+(y+)=du'^2/dy+$. Then, an interesting pattern emerges, as viewable in Figure 4. Note that the differentiation for $\varepsilon+$ is with respect to $y+$ coordinate, in contrast to $d/dy$ operation in Eq. 5. The $\varepsilon+(y+)$ curves appear to merge at all Reynolds numbers except that the positive and negative peaks progressively protrude in an inverted manner. That is, the positive peaks in $\varepsilon+(y)$ increases with $Re_\tau$, while the negative peak decreases in magnitude. Therefore, the dissipation curves are self-similar with an inverted proportionality on either side of the zero crossing. This calls for an asymmetrical scaling for the front ($y+ < y+_{peak}$) and back ($y+ > y+_{peak}$) sides, as depicted in Figures 5(a) and (b). For the front side, the scaling factor increases while for the back side an opposite is applicable. Multiplying these asymmetrical scaling factors to a template profile at a very low Reynolds number ($Re_\tau =110$) replicates the $du'^2/dy+$ curves at all other up to $Re_\tau =5200$. This is a vindication that the dissipative structure scales with the Reynolds number.

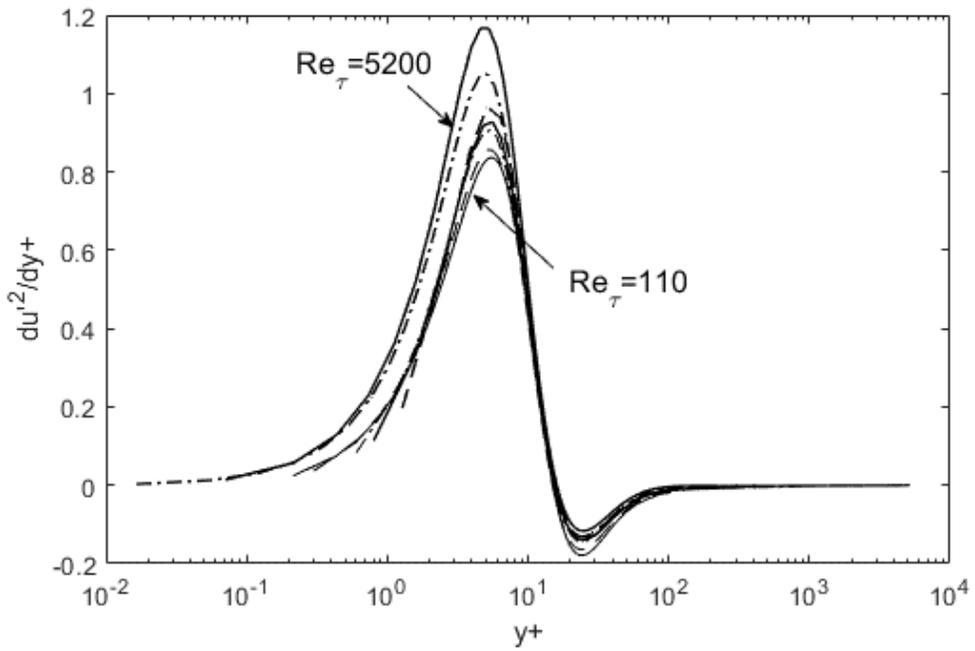

**Figure 4.** The dissipation scaling of du'²/dy+. The near-wall peaks ascend, and the negative undulations reduce in magnitude with increasing Reynolds number. DNS data of Iwamoto et al. (2002) and Graham et al. (2016) are used.

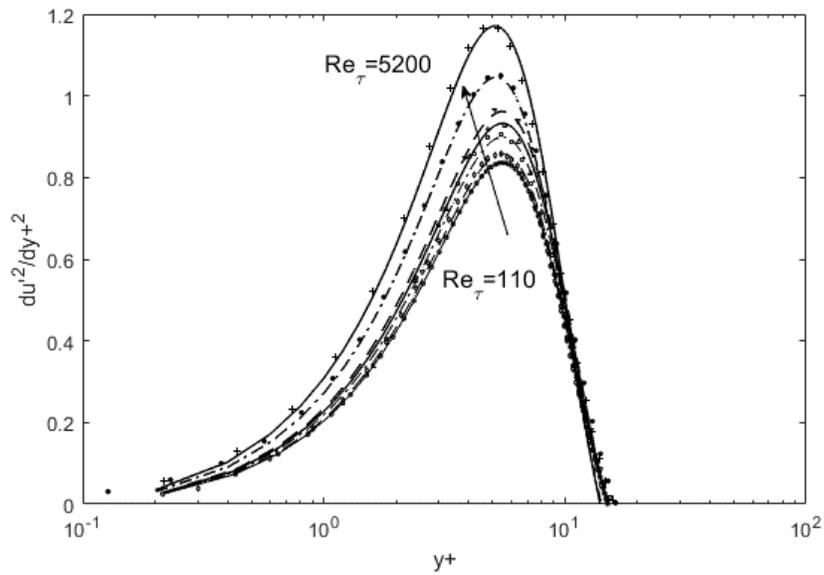

(a)

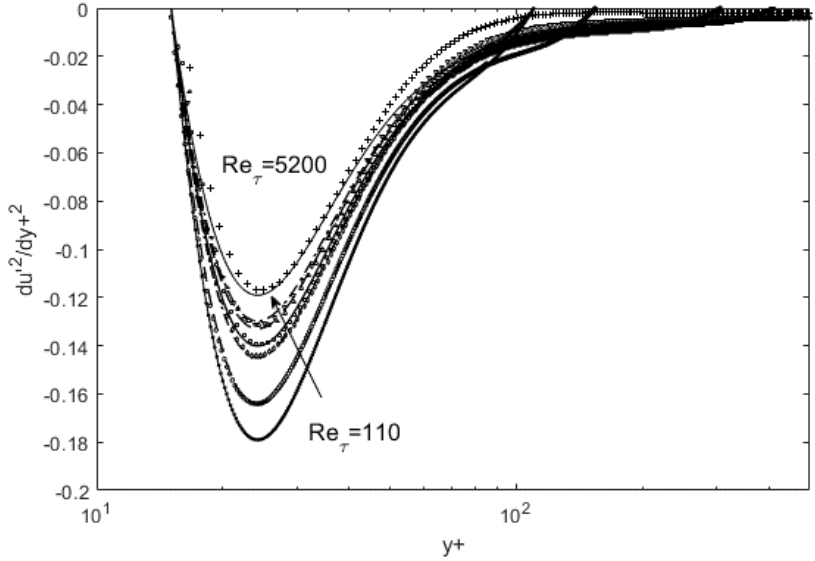

(b)

**Figure 5. Comparison of the scaled profiles based on a template at Re$_\tau$=110, with DNS data (Iwamoto et al., 2002; Graham et al., 2016). (a) Near-wall side, prior to the zero-crossing; and (b) aft side.**

With asymmetrical boundary conditions (u'$^2$ = 0 at y=0 and u'$^2$ > 0 at the centerline), internal transport dynamics (as prescribed by Eq. 3) stipulates the resulting u'$^2$ distributions in space, while encumbered by the global constraints of E=const. and Φ ~ Re$_\tau$. As shown earlier in Figure 1, the interior fluxes are predicated by Eq. 3, and the dissipation scaling exhibited in Figures 4 and 5 is the remnant of this re-distribution process. Therefore, the self-similarity in du'$^2$/dy+ can be considered as a manifestation of the Reynolds number being the sole dynamical parameter in the transport equation such as Eq. 3. Regardless of whether one agrees with this interpretation, an irresistible characteristic of this scaling is that we can start from base template of the du'$^2$/dy+ profile at any (preferably low) Reynolds number and reconstruct the curves at other (higher) Re$_\tau$. Although at high Reynolds numbers the curves start running out of y+ range and also some deviations are seen at high Reynolds number on the back side (Figure 5b), the scaling covers a wide dynamic range (Re$_\tau$ = 110 to 5200) and across two data sets (Iwamoto et al., 2002; Graham et al., 2016) separated by authors, computational methods, and time. The errors are mostly on the backside and within tolerable margins, and can be remedied by using a baseline template at a higher Reynolds number and/or using the centerline boundary conditions in corrective extrapolation algorithms. Thus, the du'$^2$/dy+ profiles exude signs of universal self-similarity. With this kind of scaling and via the transport equations (Eqs. 1-3), now a possibility is presented for reconstruction of the entire turbulence structure, as demonstrated in the next section.

A cursory examination of the v'$^2$ profiles suggests that quite different dynamics may be present. The first gradients reveal a more intelligible pattern, but not to the point of being scalable. In Figure 6(a), we can see that the peaks of the dv'$^2$/dy+ elevates in a similar manner as the u'$^2$ profiles, but the trailing edges continue to digress outward. On the other hand, the aligned peaks indicate that the zero-crossing points in the second gradients should all merge. Thus if we venture one step further and take the second gradients (d$^2$v'$^2$/dy+$^2$), a hidden structure is bared again (Figures 6b and 7) that exhibits self-similarity for nearly two orders of magnitude in the Reynolds number range, Re$_\tau$ =110 to 5200. The scaling factors this time monotonically increase on both sides of the zero-crossing point. The governing equations for u'$^2$ and v'$^2$ are evidently different: Eqs. 2 and 3 are momentum- and energy-conserving equations (Lee, 2021), respectively. A conjecture may be offered for this higher-order scaling: energy variables is dissipated according to the first gradient squared while the momentum diffusion is prescribed by the second gradient, thus resulting in sequential gradient scaling of the energy and momentum, respectively. Reconstructions of the d$^2$v'$^2$/dy+$^2$ profiles simply involve multiplications by monotonically increasing scaling factors, that are disproportionate for the front and back sides. A second-gradient template at a lower Reynolds number (e.g. Re$_\tau$=300) can be expanded through a simple arithmetic stretching to replicate the second gradients of v′² at any other Reynolds number (Re_τ=1000 and 5200), as shown in Figure 7. Due to the relatively benign slopes involved in v′² profiles, in comparison to u′², the agreement between the synthesized and DNS is almost irreproachable.

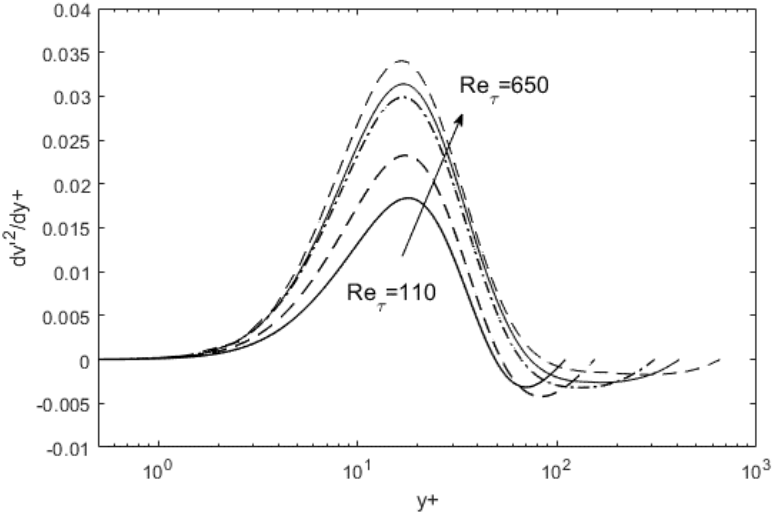

(a)

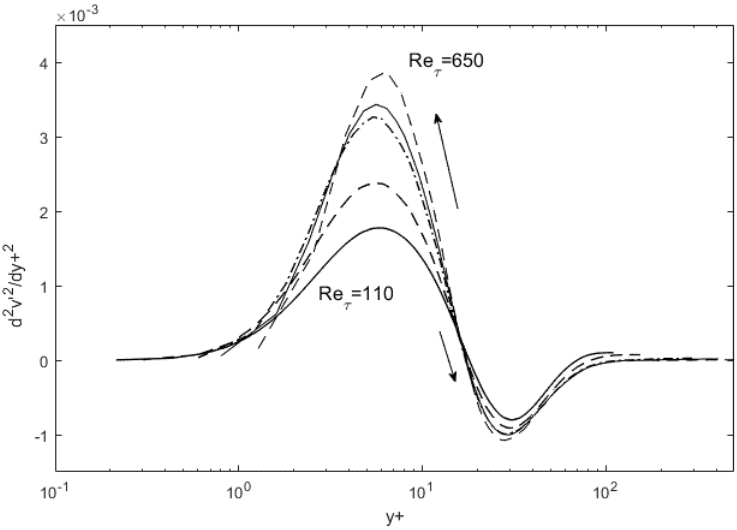

(b)

**Figure 6. (a) Progression of dv′²/dy+; and (b) dissipation scaling for v′² profiles involves the second gradients.**

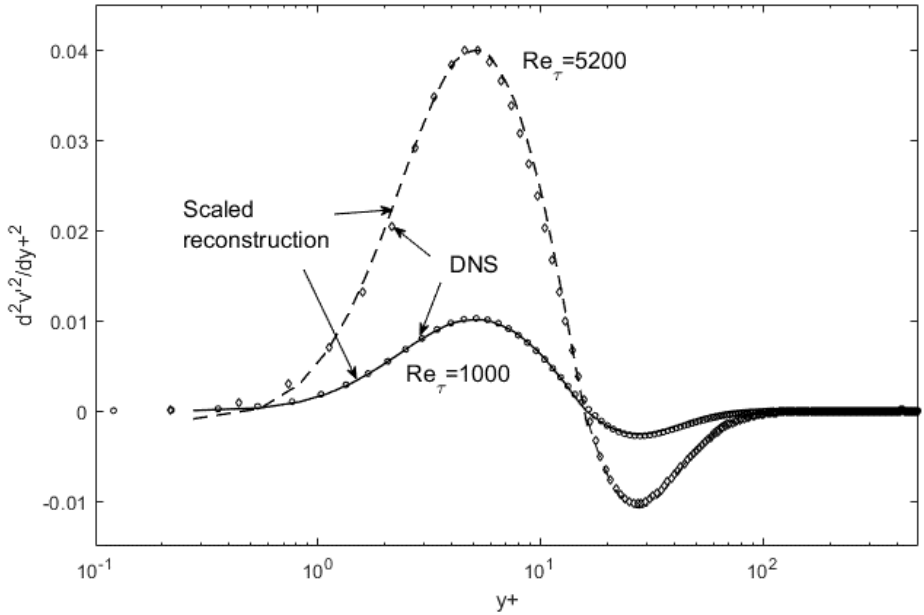

**Figure 7. Reconstruction of the full $dv'^2/dy+^2$ profiles, using a template at $Re_\tau=300$.**

Now then, what about the Reynolds shear stress itself, $u'v'$? The profiles for the $d^2u'v'/dy+^2$ appear progressively amplified in Figure 8, but embody some deviatoric behavior. Zero-crossing points tend to merge, but there is a contour change as the Reynolds number increases, where both the negative- and positive-segment widths broaden (y+ coordinate is no longer of universal utility). Also, the scaling factor exhibits a steady increase for $Re_\tau=110$ to 650, but jumps and varies by a small amount from $Re_\tau$ =1000 to 5200. A scaling behavior is not as transparent for the Reynolds shear stress at this point. A possible explanation is that $u'^2$ and $v'^2$ flux terms scale differently, and the cross-transport ($u'v'$) contains a mixture of the contributing elements, resulting in scaling deviations. y+ and $u_\tau$ (or $u_\tau^2$) are global scaling parameters based on the total momentum (or energy) available. Self-similarity arises based on these normalization parameters, but the same set of operations does not seem capable of scaling the intra-transport dynamics. There may yet be some corrective operations or alternative coordinates utilizable specifically for a "mixed" variable such as the Reynolds shear stress, but we leave open such a possibility and its search. Instead, we already acquisitioned self-similar, scalable u'$^2$ and v'$^2$ gradients, and thus can now call upon the transport equation (Eq. 1) to figure out the Reynolds shear stress variations by using them as inputs, as shall be exemplified in the next section. But before that, above findings on the dissipation and energy scaling of turbulence structure in channel flows may be summarized as follows.

**Observation 1: The integrated turbulence kinetic energy (Eq. 4) remains constant when normalized by $u_\tau^2$.**

**Observation 2: The integrated dissipation (as defined by Eq. 5) increases linearly with $Re_\tau$.**

**Observation 3: The dissipation structure for u'$^{2+}$ is prescribed by its gradient in the y+ space, scalable on the front (y+<$y_{peak}$+) and back (y+>$y_{peak}$+) sides with inverted proportionality.**

**Observation 4: The dissipation structure for v'$^{2+}$ is prescribed by its second gradient in the y+ space, scalable on the front (y+<$y_o$+) and back (y+>$y_o$+) sides disproportionately.**

**Observation 5: The contours of u'v' offers a remote possibility for self-similarity through further manipulations, but are not scaled at this point.**

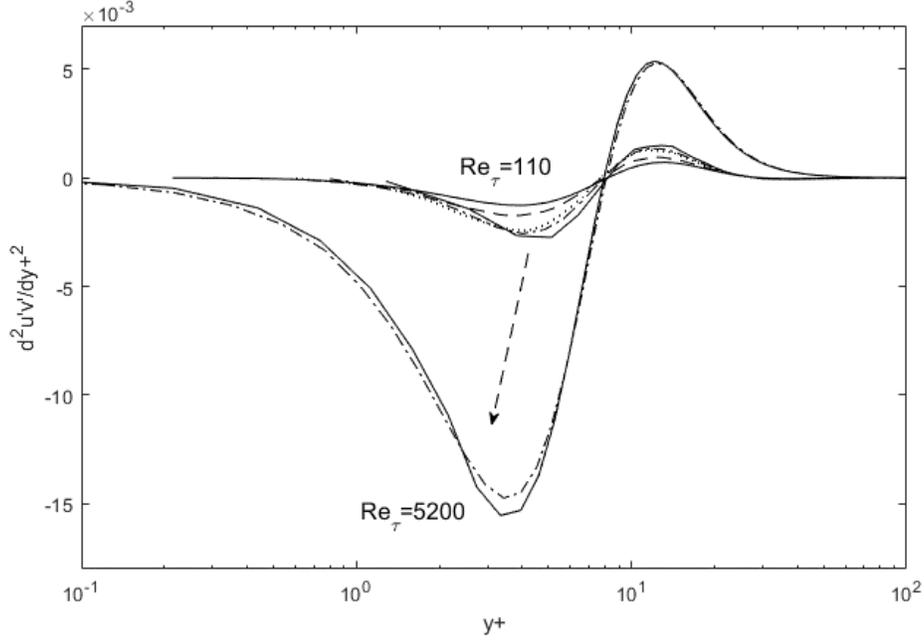

**Figure 8.** The progression of the second gradients of u'v'. DNS data of Iwamoto et al. (2002) and Graham et al. (2016) are used.

**RECONSTRUCTION OF THE TURBULENCE STRUCTURE**

The observations of the dissipation scaling, along with transport equations (Eqs. 1-3), direct us to a viable route for reconstructing the complete structures of turbulence channel flows, including the mean velocity profile (via the Reynolds shear stress supplanted in RANS). Let us suppose that from the scaling operations shown above (or through an iterative algorithm to solve Eqs. 1-3), we have reconstructed $u'^2$ and $v'^2$ profiles with a fair degree of accuracy at a given Reynolds number. Then, their effects on the Reynolds shear stress, u'v', can be computed using Eq. 1, at any other Reynolds numbers. As an example, using the template at a lower Reynolds number ($Re_\tau$=650), the scaling factors used earlier are ratioed to reconstruct $u'^2$ profiles at $Re_\tau$=1000 and 5200, plotted as solid lines in Figure 9. Iterative solutions of Eqs. 1-3 (Lee, 2021) are also included for comparison. Both approaches result in good agreement with DNS data near the wall, but tend to misalign on the descending side. Primary reason is that y+ coordinate taken from a lower Reynolds number does not cover above y+ ~1000. Also, the reconstructions are the result of numerically integrating $du'^2/dy+$, so that any minute errors in the gradients will lead to mis-tracking of the profile. However, on the positive side it is the gradients of $u'^2$ and $v'^2$ that go into determination of the $u'v'$ in Eq. 1, and also near-wall precision is acceptable in Figure 9, which is far more requisite as most of the turbulent "actions" occur in that region. In addition, the accuracy and coverage of the descending segment(s) can be correctively improved using the centerline boundary conditions as secondary constraints. Profiles for $v'^2$ fare significantly better in all of these regards, as can be anticipated from the faithful tracking exhibited earlier in Figure 7.

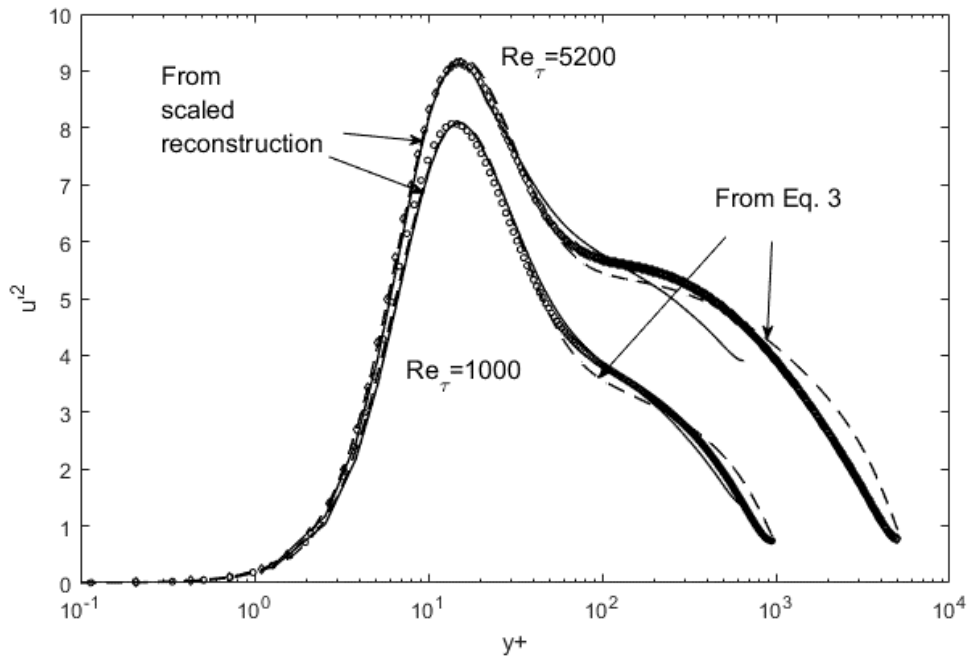

**Figure 9. Comparison of reconstructed $u'^2$ profiles with DNS data of Graham et al. (2016).**

Once we have the reconstructed u'² and v'² gradients, then these are all the necessary ingredients to compute u'v' using Eq. 1. The gradient d(u'v')/dy on the left-hand side of Eq. 1 is computed, for subsequent numerical integration for direct comparison with DNS data again. To reduce errors, integration is started from both the wall and the centerline, until the marched solutions overlap or intersect. The outcome of this process is illustrated in Figure 10. There are deviations from y+ ~30 to 50, but overall the reproducibility of the Reynolds shear stress is creditable, validating the succinct momentum dynamics embodied in Eq. 1 (Lee, 2020; 2021).

Thus, the Reynolds number effects on the turbulence structure in channel flows have origins in the u'² structure, constrained to conserve energy while maximizing the dissipation to the limit imposed by the viscosity. As we saw, this demands higher skewness, and the way that a continuous and smooth function achieves this is through a sharpening peak moving closer to the wall with increasing Reynolds number. The u'² profiles, as governed by Eq. 3, are scalable according to its fluxes. The gradients of u'², and v'² (the pressure work term in Eq. 3) then prescribe d(u'v')/dy according to the momentum dynamics contained in Eq. 1 (Lee, 2020). Therefore, we can visualize the Reynolds number effects as manifestations of energy-conserving and dissipative restructuring processes of u'² and v'², along with subsequent transport or redistribution dynamics.

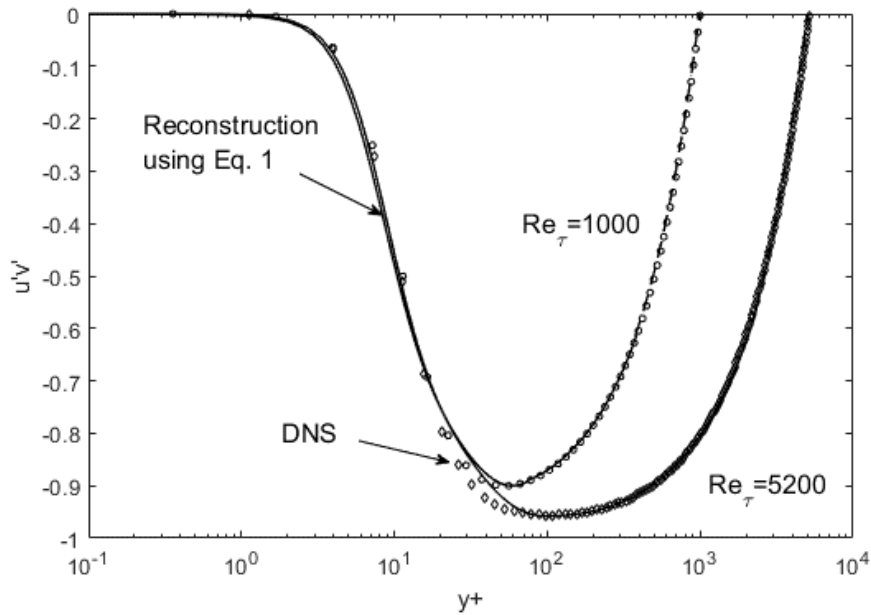

**Figure 10. The Reynolds shear stress computed using Eq. 1, based on reconstructed $u'^2$ and $v'^2$ profiles.**

The final destination of the above sequence of events is the mean velocity structure, which simply follows the integral of the Reynolds shear stress according to the mean momentum or RANS equation(s), as shown in Figure 11. We can see that any errors present in u'v' near y+ ~ 30 again translate to the mean velocity calculations. Otherwise, the usual features of the mean momentum distribution are all observable in Figure 11, such as the viscous layer with constant slope near the wall, transitioning to a gradual slope toward the centerline boundary condition of dU/dy=0. Any effects of the Reynolds number, scalable or otherwise, on the mean velocities would emerge from the u'v' structure, and as noted above it is an open exercise to seek any self-similarities or patterns in the Reynolds shear stress or its gradient profiles. Regardless of future findings on the u'v' scalability, a viable method exists at this juncture for reconstructing the mean velocity profiles at an arbitrary Reynolds number, through the dissipation scaling of u′² and v′² and the transport dynamics contained in Eqs. 1-3.

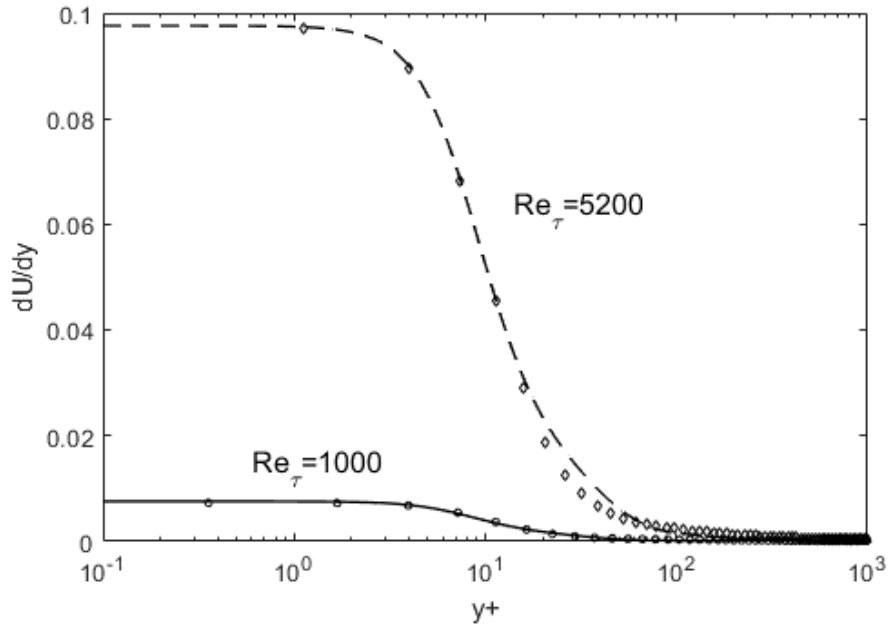

**Figure 11. Input of u′v′ in the once-integrated RANS results in the mean velocity gradients.**

**CONCLUDING REMARKS**

From the discussions above, we can see that the initiating effect of the Reynolds number is to increase the maximum allowable total dissipation, Φ, while u′² and v′² must contain a fixed total kinetic energy, as normalized by the energy scale, $u_\tau^2$. The route in which Φ elevates is by skewing the spatial distribution, i.e. (du′²/dy) must increase across the board. However, u′² and v′² transports are internally constrained by the flux terms in Eqs. 2 and 3, and also must follow continuous and smooth functional progressions. This interior dynamics via gradient transport apparently causes self-similarity in these variables. The fact that $u'^2$ and $v'^2$ transports are dictated by energy and momentum fluxes, respectively, is a potential cause of the dissipation following the first and second gradients accordingly. This ordered structure allows for reconstructive operations at any Reynolds numbers, based on a template. The gradients of $u'^2$ and $v'^2$ are key terms in the Reynolds shear stress expression (Eq. 3), so that they lead to computed $u'v'$ distributions, which in turn generate the mean velocity profiles, both within reasonable margins of error relative to the DNS data. These scaling characteristics along with the transport equations (Eqs. 1-3) furnish succinct insights and computability for the interior dynamics of turbulent channel flows.